\PassOptionsToPackage{unicode}{hyperref}
\PassOptionsToPackage{hyphens}{url}
\PassOptionsToPackage{dvipsnames,svgnames*,x11names*}{xcolor}
\documentclass[
  11pt]{article}
\usepackage{lmodern}
\usepackage{amssymb,amsmath}
\usepackage{ifxetex,ifluatex}
\ifnum 0\ifxetex 1\fi\ifluatex 1\fi=0 
  \usepackage[T1]{fontenc}
  \usepackage[utf8]{inputenc}
  \usepackage{textcomp} 
\else 
  \usepackage{unicode-math}
  \defaultfontfeatures{Scale=MatchLowercase}
  \defaultfontfeatures[\rmfamily]{Ligatures=TeX,Scale=1}
\fi
\IfFileExists{upquote.sty}{\usepackage{upquote}}{}
\IfFileExists{microtype.sty}{
  \usepackage[]{microtype}
  \UseMicrotypeSet[protrusion]{basicmath} 
}{}
\makeatletter
\@ifundefined{KOMAClassName}{
  \IfFileExists{parskip.sty}{%
    \usepackage{parskip}
  }{
    \setlength{\parindent}{0pt}
    \setlength{\parskip}{6pt plus 2pt minus 1pt}}
}{
  \KOMAoptions{parskip=half}}
\makeatother
\usepackage{xcolor}
\IfFileExists{xurl.sty}{\usepackage{xurl}}{} 
\IfFileExists{bookmark.sty}{\usepackage{bookmark}}{\usepackage{hyperref}}
\hypersetup{
  pdftitle={How many labels can a biological oscillator carry? A quality-factor screen for proposed information carriers},
  pdfauthor={Eran Kopel},
  colorlinks=true,
  linkcolor=blue,
  filecolor=Maroon,
  citecolor=Blue,
  urlcolor=blue,
  pdfcreator={LaTeX via pandoc}}
\usepackage[margin=2.5cm]{geometry}
\usepackage{longtable,booktabs}
\usepackage{etoolbox}
\makeatletter
\patchcmd\longtable{\par}{\if@noskipsec\mbox{}\fi\par}{}{}
\makeatother
\IfFileExists{footnotehyper.sty}{\usepackage{footnotehyper}}{\usepackage{footnote}}
\makesavenoteenv{longtable}
\providecommand{\tightlist}{%
  \setlength{\itemsep}{0pt}\setlength{\parskip}{0pt}}
\usepackage[T1]{fontenc}
\usepackage{lmodern}
\usepackage{amsmath,amssymb}
\usepackage{textcomp}
\usepackage{booktabs}
\usepackage{longtable}
\usepackage{microtype}
\usepackage{caption}
\usepackage{url}
\usepackage{etoolbox}
\AtBeginEnvironment{longtable}{\footnotesize}
\title{How many labels can a biological oscillator carry? A
quality-factor screen for proposed information carriers}
\author{Eran
Kopel\thanks{Tel Aviv University, Tel Aviv, Israel. \texttt{erankopel@tauex.tau.ac.il}, ORCID \texttt{0000-0003-4657-8636}.}}
\date{}

\begin{document}
\maketitle

\hypertarget{keywords}{%
\section*{Keywords}\label{keywords}}
\addcontentsline{toc}{section}{Keywords}

Neural oscillations; Information capacity; Quality factor; Gamma rhythm;
Cortical minicolumn; Quantum biology

\begin{center}\rule{0.5\linewidth}{0.5pt}\end{center}

\hypertarget{abstract}{%
\section*{Abstract}\label{abstract}}
\addcontentsline{toc}{section}{Abstract}

How many distinguishable labels can a biological oscillator carry?
Proposals invoking collective vibrational modes, endogenous
electromagnetic fields, microtubule excitations and oscillatory phase
codes are each debated on grounds particular to themselves, with no
shared standard for comparison. We show that spectral distinguishability
alone bounds the number of labels by the quality factor,
\(M \le Q = 2\pi\nu\tau_{\mathrm{coh}}\). This follows from the relation
between linewidth and coherence time, so it is independent of substrate,
of mechanism, and of any position on quantum effects in biology, and it
can be evaluated from two published quantities. Applied to a recently
proposed 30 GHz intracolumnar microwave field in cortex, it gives
\(Q = 0.19\): the linewidth exceeds the carrier fivefold. The obvious
rescue, that a driven emitter can be spectrally narrower than its gain
medium, requires a resonant cavity, and the model's own geometry forbids
one. An independent bound on metabolic power is exceeded by five to nine
orders of magnitude. Six further criteria follow from the same
standpoint, including a two-sided persistence window requiring a label
to be both readable and rewritable. Screening eleven carriers, only the
low-frequency neural rhythms pass. High-frequency molecular carriers are
eliminated by brevity, not by the fragility the debate has assumed.

\begin{center}\rule{0.5\linewidth}{0.5pt}\end{center}

\hypertarget{introduction}{%
\section{Introduction}\label{introduction}}

Arguments about physical carriers of information in biological tissue
have a characteristic failure mode. A mechanism is proposed. Critics
argue that it is implausible on thermal or decoherence grounds (Tegmark,
2000). Proponents reply that the protection mechanism has been
underestimated. The exchange does not converge, and it does not converge
for a structural reason: both sides are arguing about whether the
mechanism can \emph{exist}, and existence claims in this domain are
difficult to settle, since they turn on contested estimates of coupling
strengths, screening, and the applicability of open-system treatments to
warm and wet matter.

The more tractable question is whether the mechanism could
\emph{function}. A carrier that exists but cannot support
distinguishable states, or cannot be read within the time available, or
cannot be rewritten, or cannot be paid for out of the local energy
budget, is of no use whatever its physical status. Function-level
requirements have three properties that make them worth pursuing. They
are quantitative. They are largely substrate-independent, since they
follow from what a label has to do rather than from what it is made of.
And they can be evaluated without settling the contested physics, which
means an argument built from them cannot be answered by relitigating
decoherence.

The absence of such a standard carries a cost on both sides. A physicist
proposing a mechanism has no agreed target to aim at, and so aims at
existence, which is at once the hardest property to establish and the
least informative one. A biologist assessing the proposal has no cheap
way to test it, and so either accepts the physics on authority or
dismisses the whole class on principle. A screen that either party can
apply in an afternoon, from quantities both already publish, would let
the exchange proceed without either having to concede the other's ground
in advance.

\hypertarget{the-biological-problem-the-criteria-address}{%
\subsection{The biological problem the criteria
address}\label{the-biological-problem-the-criteria-address}}

The question is not abstract. Several standing problems in systems
neuroscience would be solved, or at least reframed, if cortical tissue
possessed a mechanism for tagging activity with its origin. Accounts of
perceptual binding require some means of marking which features belong
to which object; accounts of source monitoring require a means of
marking which activity is internally generated. Both traditions have had
to postulate a tag rather than derive one, and a physical mechanism
supplying column-specific identity would be of immediate use to them.

Cortical gamma is the best-characterised candidate, and it is well
enough characterised for the criteria below to be applied to it rather
than merely asserted of it. Its frequency and reliability have been
measured directly (Tan et al., 2016), its synchronisation dynamics have
a quantitative theory (Lowet et al., 2017), and, importantly for the
writability criterion developed in §3, its peak frequency appears to be
under neurochemical rather than structural control: resting GABA
concentration was reported to predict individual gamma frequency
(Muthukumaraswamy et al., 2009), though a larger study did not replicate
the association (Cousijn et al., 2014). A carrier whose frequency can be
set by a diffusible signal is a candidate label; one whose frequency is
fixed by immutable molecular properties is not.

The framework below is intended to let such candidates be compared on
common terms, and to let proposals that cannot serve the function be
identified cheaply.

\hypertarget{the-bound}{%
\subsection{The bound}\label{the-bound}}

Let a \emph{carrier} be a physical degree of freedom whose state is
intended to convey which of \(M\) alternatives obtains: a label, an
address, or a channel index. Suppose the alternatives are encoded in the
carrier's frequency, which is the encoding assumed whenever a mechanism
is described as supplying a ``specific frequency'' to a structure.

Two instances of the carrier are distinguishable only if their
frequencies differ by more than the linewidth. For a mode with coherence
time \(\tau_{\mathrm{coh}}\) the linewidth is

\[\delta\nu \;\approx\; \frac{1}{2\pi\tau_{\mathrm{coh}}},\]

so the number of resolvable values that can be packed into a band of
width \(B\) is \(M \le B/\delta\nu\). Taking the band to scale with the
carrier itself, \(B \sim \nu\), which is the most favourable assumption
available to any such proposal, gives

\begin{equation}
M \;\le\; 2\pi\nu\,\tau_{\mathrm{coh}} \;=\; Q.\tag{1}
\end{equation}

The number of distinguishable frequency labels a carrier can support is
therefore bounded by its quality factor. Nothing in the derivation
refers to the physical nature of the oscillator. It applies equally to a
molecular vibration, a collective electromagnetic mode, and a population
rhythm, and it requires exactly two published quantities to evaluate.

The degenerate case is worth stating separately, because it is the one
that arises in practice. A necessary condition is \(Q > 1\). A system
with \(Q < 1\) is not an oscillator at all. Its excitation decays before
a single cycle completes, its ``frequency'' is not a property that can
differ reproducibly between instances, and there is consequently nothing
available to be labelled by.

Three scope limits. First, Eq. (1) governs \emph{frequency-multiplexed}
codes. Carriers that encode in amplitude, in relative timing, or in
chemical identity are bounded differently, and we return in §4.1 to why
biological systems overwhelmingly use the latter. Second, the bound
concerns the source. Whether a downstream reader can resolve \(M\)
values is a separate question, governed by an estimation-theoretic floor
developed as criterion C2 in §3, which turns out not to bind when \(Q\)
is satisfied.

Third, and least obvious, Eq. (1) treats the linewidth as
\emph{homogeneous}, that is, as arising from dephasing rather than from
slow wandering of the centre frequency. The two broaden a time-averaged
spectrum identically but differ in what a reader can recover. Where a
measured width is dominated by inhomogeneous broadening, a decoder that
tracks the drift within its observation window resolves more values than
\(\delta\nu\) implies, and \(Q\) read off an averaged spectrum
understates \(M\). This does not affect §2, where the coherence time is
a vibrational dephasing time obtained by two-dimensional infrared
spectroscopy, a technique that separates the two contributions by
construction, and where a carrier with \(Q<1\) completes no cycle for a
decoder to track. It does affect the physiological entries of Table 3,
whose coherence times are inferred rather than resolved into homogeneous
and inhomogeneous parts, and it sharpens the requirement of §6.3: what
must be measured is the damping, not the width of an averaged spectrum.

\hypertarget{plan}{%
\subsection{Plan}\label{plan}}

The bound is elementary, and we make no claim otherwise. Its interest
lies in what it excludes. §2 applies it to a specific, quantitative,
recently published proposal that a 30 GHz intracolumnar microwave field
serves as a column-specific recognition signal in cortex (Keppler, 2023,
2025). The proposal fails the bound by a wide margin, survives neither
of the two available rescues, and fails a second and independent bound
on power. §3 then asks what a carrier needs beyond distinguishability,
and obtains six further criteria from the same standpoint. §4 assembles
them into a screen and applies it across the candidate carriers
currently under discussion. §5 states what a proposal would have to
supply in order to pass. §§6 and 7 discuss what the analysis does and
does not establish, and where the framework's own limits lie. An earlier
and partial statement of the framework of §§3 to 5, without the case
analysis, was deposited as Kopel (2026a).

\begin{center}\rule{0.5\linewidth}{0.5pt}\end{center}

\hypertarget{a-worked-rejection-the-30-ghz-intracolumnar-field}{%
\section{A worked rejection: the 30 GHz intracolumnar
field}\label{a-worked-rejection-the-30-ghz-intracolumnar-field}}

\hypertarget{the-proposal-and-the-reproduction-of-its-arithmetic}{%
\subsection{The proposal, and the reproduction of its
arithmetic}\label{the-proposal-and-the-reproduction-of-its-arithmetic}}

A recent quantum-electrodynamic model of the cortical microcolumn
(Keppler, 2023, 2025), built on the coherent-QED formalism of Preparata
(1995) and the coherent-domain programme of Del Giudice et al.~(2005),
proposes that the glutamate pool couples resonantly to the
electromagnetic zero-point field and forms a coherence domain of order
\(10^{11}\) molecules and 30 \textmu{}m in diameter, generating an intracolumnar
microwave field near 30 GHz. The field is advanced as a column-specific
recognition signal: a physical tag by which one cortical column may be
distinguished from another (Keppler, 2024, 2025).

The proposal is worth engaging on its own terms for two reasons. It is
quantitative, and therefore checkable, which is uncommon in this
literature. And the claim it makes is one that the traditions described
in §1.1 have independent use for.

We begin with what survives. The model's internal arithmetic reproduces.
The coupling scales as \(\sqrt{n}\) with concentration as claimed, with
\(\sqrt{12/300} = 0.200\) at tissue against vesicular concentration; the
coherence-domain diameter matches
\(\tfrac{\pi}{4}\lambda(7.8\ \mathrm{THz})\) to within 0.6\%; and the
molecule count recomputes as \(1.02\times10^{11}\) against a quoted
\(10^{11}\). The full reproduction is given in Appendix A. Internal
consistency is not in question, and the difficulties we identify lie
elsewhere.

\hypertarget{the-linewidth}{%
\subsection{The linewidth}\label{the-linewidth}}

The proposed function is labelling. By Eq. (1) that requires
\(Q = \nu_c/\delta\nu \gg 1\), which requires in turn a coherence time
for the mode.

Direct dephasing measurements for glutamate at 260 cm\(^{-1}\) in
aqueous solution do not appear to exist. We therefore take values from
femtosecond two-dimensional infrared and pump-probe studies of
carboxylate modes in aqueous solution (Kuroda et al., 2010; Kuroda and
Hochstrasser, 2011, 2012; Korotkevich and Bakker, 2022): the right
molecular group, the right technique, and the right solvent, though
measured at the carboxylate stretch near 1400 to 1600 cm\(^{-1}\) rather
than at 260 cm\(^{-1}\). These give \(T_2\) of order 1 ps. Table 1
evaluates \(Q\) across the resulting range.

\begin{longtable}[]{@{}lllll@{}}
\caption{\textbf{Table 1.} Quality factor of a 30 GHz carrier at
candidate coherence times.}\tabularnewline
\toprule
\(\tau_{\mathrm{coh}}\) & basis & \(\delta\nu\) (Hz) & \(Q\) at 30 GHz &
label possible\tabularnewline
\midrule
\endfirsthead
\toprule
\(\tau_{\mathrm{coh}}\) & basis & \(\delta\nu\) (Hz) & \(Q\) at 30 GHz &
label possible\tabularnewline
\midrule
\endhead
1 ps & carboxylate 2D-IR, generous proxy & \(1.6\times10^{11}\) &
\textbf{0.19} & no\tabularnewline
27 fs & water far-infrared continuum & \(5.9\times10^{12}\) &
\textbf{0.005} & no\tabularnewline
1 ns & \(10^{3}\times\) enhancement over proxy & \(1.6\times10^{8}\) &
188 & yes\tabularnewline
1 ms & as the model requires & 159 & \(1.9\times10^{8}\) &
yes\tabularnewline
\bottomrule
\end{longtable}

At the generous proxy the linewidth is five times the carrier frequency.
Every column's frequency overlaps every other column's, and \(Q < 1\)
places the domain outside the class of objects that have a frequency in
the required sense.

\hypertarget{which-way-the-substitution-errs}{%
\subsection{Which way the substitution
errs}\label{which-way-the-substitution-errs}}

The proxy is not a measurement of the mode in question, and it is worth
being explicit about the direction of the error, because it runs in the
model's favour rather than ours.

A mid-infrared stretch near 1500 cm\(^{-1}\) must relax through
multi-quantum processes, and water offers relatively few resonant
accepting modes at that frequency, which is why the measured timescales
are picoseconds. The mode the model requires sits at 260 cm\(^{-1}\),
inside water's own intermolecular band: hindered translations near 50 to
200 cm\(^{-1}\) and the rising edge of the librational band above 400
cm\(^{-1}\). A solute mode there is directly resonant with the solvent's
collective motions, in a spectral region where water absorbs as a broad
and essentially featureless continuum. A mode embedded in a continuum of
width \(\Delta\tilde\nu\) inherits a correlation time of order
\(1/2\pi c\,\Delta\tilde\nu\), which is 13 to 53 fs for widths of 400 to
100 cm\(^{-1}\).

The appropriate estimate at 260 cm\(^{-1}\) is therefore tens of
femtoseconds rather than picoseconds. The picosecond figure is an upper
bound we grant, not a measurement we rely on, and the true margin on the
conclusion is four to forty times larger than the headline \(Q = 0.19\)
suggests.

\hypertarget{the-rescue-and-why-it-is-unavailable}{%
\subsection{The rescue, and why it is
unavailable}\label{the-rescue-and-why-it-is-unavailable}}

A reader should object at this point, and the objection is correct in
general:

\begin{quote}
A laser's linewidth is far narrower than the dephasing rate of its gain
medium. Setting \(\delta\nu = 1/2\pi T_2\) is therefore not valid for a
\emph{driven} oscillator. If the intracolumnar field is a driven
collective emission rather than a bare molecular coherence, its
linewidth could be orders of magnitude narrower than \(1/T_2\), and
\(Q = 0.19\) does not follow.
\end{quote}

This is right as stated. Schawlow-Townes narrowing is real, and it is
why a laser can be spectrally pure despite a broad gain medium. But
narrowing requires a resonant structure. The question is therefore
whether a coherence domain is a cavity at 30 GHz, and the answer follows
from the model's own geometry (Appendix A, Table A.2).

The domain diameter is \emph{derived} as
\(\tfrac{\pi}{4}\lambda(7.8\ \mathrm{THz}) = 30\) \textmu{}m. The structure is
thereby defined to be resonant at 7.8 THz. The carrier is then said to
shift to 30 GHz, where the same structure is \(1/333\) of a wavelength
across. A structure cannot resonate at a frequency whose wavelength is
333 times its own size. The 30 \textmu{}m domain is a resonator near 10 THz and
at no lower frequency, so there is no cavity at 30 GHz, no narrowing
mechanism, and the rescue is unavailable.

Two features make this argument the strongest available against the
proposal. It is pure Fourier analysis together with a length comparison,
so it entails no commitment about coherent quantum electrodynamics,
about decoherence mechanisms, or about quantum effects in biology
generally, and cannot be answered by disputing any of them. And it is a
requirement for the proposed \emph{function} rather than for the
proposed physics: even granting the coherence domain in some form, it
cannot serve as a label unless \(\tau_{\mathrm{coh}} \gtrsim 30\) ps
merely to reach \(Q > 1\), and \(\gtrsim 1\) \textmu{}s to carry a useful number
of channels.

A scope limit deserves emphasis. At \(ka = 0.0094\) the domain is a
deeply sub-wavelength radiator, with far-field dipole radiation
efficiency of order \((ka)^2 \approx 9\times10^{-5}\). It would be easy
to deploy this as a further objection, that the field cannot radiate at
all. That would be overreach. The model's proposed action is on ion
channels within the same column, which is near-field coupling, and near
fields do not care about radiation resistance. The electrically-small
argument refutes the cavity rescue and nothing else. We flag the
distinction because it is easy to blur, and because the criticism is
stronger for being narrow.

\hypertarget{an-internal-tension}{%
\subsection{An internal tension}\label{an-internal-tension}}

The model itself supplies the relevant physics, in a different context.
Arguing that terahertz brain stimulation is infeasible, it observes that
such radiation, through its strong interaction with water, penetrates
only a few hundred micrometres into tissue (Keppler, 2025). That is the
same interaction. Strong water absorption near 7.8 THz entails both that
tissue is opaque there, which the model uses, and that a mode at that
frequency is heavily damped, which it does not address.

The tension is sharper than a simple oversight, because the model
\emph{requires} strong coupling to the water matrix: hydration-enhanced
dipole moments are what lift glutamate's coupling constant into the
critical regime in the first place. The coupling term that supplies the
oscillator strength is the coupling term that destroys the coherence.
These are not separable effects to be traded against one another. They
are the same interaction, and a treatment that invokes one must account
for the other.

\hypertarget{the-power-budget-and-a-circularity}{%
\subsection{The power budget, and a
circularity}\label{the-power-budget-and-a-circularity}}

A second bound, independent of the first and requiring no model of
decoherence at all. Sustaining any driven coherent mode requires energy
resupplied at a rate of at least the stored energy divided by the
dephasing time. Against a metabolic supply of 8 to 63 nW per
hundred-neuron minicolumn, derived independently from ATP turnover and
from cortical power divided by neuron count (Attwell and Laughlin,
2001), the requirement runs from 5.5 mW under the most generous
assumptions available (surface molecules only, small excited fraction)
to 264 W under the model's nominal ones. The deficit spans five to nine
orders of magnitude across every assumption we could construct.

Inverting the budget gives the sharper statement. The dephasing time the
metabolic supply can afford is

\begin{equation}
T_2 \;\ge\; \frac{N\sin^2\!\gamma\,\hbar\omega_0}{P_{\mathrm{metab}}} \;\approx\; 0.8\text{ to }4\ \mathrm{ms}.\tag{2}
\end{equation}

To be affordable, the state must already possess a millisecond dephasing
time, which is precisely the quantity the model was constructed to
explain. Metabolic pumping cannot supply it, so it must be intrinsic;
intrinsic protection requires an energy gap large against \(kT\); and
the per-molecule gap is 0.130 meV, or \(kT/205\), giving a Boltzmann
factor of 0.995 and no single-molecule protection whatever. The loop
does not close.

The excited fraction is the one quantity in this calculation we could
not extract from the model's published description, and it proves not to
be needed, because it does not enter independently. Rearranging,

\begin{equation}
N\sin^2\!\gamma \;\le\; \frac{P_{\mathrm{metab}}\,\tau_{\mathrm{coh}}}{\hbar\omega_0},\tag{3}
\end{equation}

whose left-hand side is simply the number of quanta the mode contains.
How that number divides between a large domain at low occupation and a
small one at high occupation is precisely what the missing parameters
would settle, and the bound is indifferent to it. With
\(\hbar\omega_0 = 32.3\) meV at 7.8 THz, the metabolic supply of an
entire minicolumn, assigned wholly to this single mode, affords 1.6 to
12 quanta at the generous picosecond proxy, and 0.04 to 0.3 quanta at
the dephasing time appropriate to 260 cm\(^{-1}\), which is less than a
single excitation. A collective state holding fewer than a dozen quanta
is not the coherence domain of \(10^{11}\) molecules the model
describes, whatever its occupation fraction proves to be.

\hypertarget{what-this-establishes}{%
\subsection{What this establishes}\label{what-this-establishes}}

The proposal fails to support a frequency label, on two independent
grounds, neither of which requires a position on quantum biology. This
is a narrower conclusion than a dismissal, and §6.1 sets out what
survives.

What the analysis also does, and what the remainder of this paper
pursues, is expose the general shape of the argument. Distinguishability
was one requirement among several, and the others can be derived from
the same standpoint.

\begin{center}\rule{0.5\linewidth}{0.5pt}\end{center}

\hypertarget{what-else-a-carrier-needs}{%
\section{What else a carrier needs}\label{what-else-a-carrier-needs}}

Distinguishability is necessary and not sufficient. A carrier must also
be readable, coupled to whatever it addresses, affordable, persistent on
the right timescale, writable, and thermodynamically consistent with the
regime its proponents claim for it. Each yields a criterion that is
quantitative, largely substrate-independent, and evaluable from a small
number of published quantities. We state them in that order, using the
microwave proposal of §2 as a running instance where it is informative.
Criterion C1 is Eq. (1).

\hypertarget{c2.-readout-the-estimation-floor}{%
\subsection{C2. Readout: the estimation
floor}\label{c2.-readout-the-estimation-floor}}

C1 constrains the source. C2 constrains the reader. Estimating the
frequency of a tone from an observation of duration \(T\) at total
signal-to-noise ratio \(\rho\) is bounded below by the Cramér-Rao
inequality (Rife and Boorstyn, 1974),

\begin{equation}
\sigma_f \;\ge\; \frac{1}{2\pi T}\sqrt{\frac{6}{\rho}},\tag{4}
\end{equation}

and a label is resolvable only if \(\sigma_f\) falls below the channel
spacing \(\nu/M\). This is not C1 restated. Averaging over an
observation window beats the linewidth once \(\rho > 6\), so a reader
with sufficient signal can in principle resolve more values than the
instantaneous linewidth would suggest. Table 2 gives the requirement for
a representative cortical carrier.

\begin{longtable}[]{@{}llllll@{}}
\caption{\textbf{Table 2.} Signal-to-noise ratio required to resolve
\(M\) channels, for a 40 Hz carrier observed for 150 ms.}\tabularnewline
\toprule
\(M\) & 10 & 25 & 50 & 100 & 200\tabularnewline
\midrule
\endfirsthead
\toprule
\(M\) & 10 & 25 & 50 & 100 & 200\tabularnewline
\midrule
\endhead
required \(\rho\) & 0.42 & 2.6 & 10.6 & 42 & 169\tabularnewline
\bottomrule
\end{longtable}

At the channel counts biology plausibly needs, the requirement is
undemanding. C2 is therefore usually not binding when C1 is satisfied,
which is worth knowing in its own right: improving signal-to-noise buys
nothing until coherence time improves. It also means that a proposal
failing C1 cannot be rescued by appeal to a sensitive reader.

\hypertarget{c3.-coupling-a-stated-mechanism-with-adequate-range}{%
\subsection{C3. Coupling: a stated mechanism with adequate
range}\label{c3.-coupling-a-stated-mechanism-with-adequate-range}}

Radiative and non-radiative carriers face different tests, and
conflating them is a common error.

A \emph{radiative} carrier requires an emitting structure of size
comparable to the wavelength, or else an explicit account of near-field
coupling with a stated range. The point that matters is not radiation
efficiency but resonance: a structure with \(d \ll \lambda\) cannot host
a mode at that wavelength, which is the argument that closed the cavity
rescue in §2.4. A \emph{quasi-static or collective} carrier, such as an
extracellular field or a chemical wave, is not constrained by
\(\lambda\), but must have a coupling range at least equal to the
distance over which the label has to be communicated.

The criterion is that a mechanism be stated and its range checked
against the communication distance. Proposals frequently leave both
implicit, and the omission conceals the difference between a carrier
that addresses its neighbours and one that addresses a region.

\hypertarget{c4.-energetic-sustainability}{%
\subsection{C4. Energetic
sustainability}\label{c4.-energetic-sustainability}}

Maintaining a carrier against dissipation costs power. For \(N\)
elements each carrying excitation \(\hbar\omega\) with excited fraction
\(f\), sustained against a coherence time \(\tau_{\mathrm{coh}}\),

\begin{equation}
P \;=\; \frac{N f \hbar\omega}{\tau_{\mathrm{coh}}} \;\le\; P_{\mathrm{metab}}.\tag{5}
\end{equation}

§2.6 evaluated this for the microwave proposal and found a deficit of
five to nine orders of magnitude. The general structure is worth
isolating from that instance, because it is counterintuitive and it
closes an escape route: shorter coherence times make a carrier
\emph{more} expensive, not cheaper, since the mode must be
re-established more often. A proposal cannot concede a low \(Q\) in
order to satisfy C1's critics and then decline to pay for the concession
in C4. The two criteria pull in opposite directions, and a candidate
must satisfy both at the same value of \(\tau_{\mathrm{coh}}\).

\hypertarget{c5.-the-persistence-window}{%
\subsection{C5. The persistence
window}\label{c5.-the-persistence-window}}

Less commonly stated, and two-sided. A label must persist long enough to
be read, and must also be rewritable fast enough to track whatever it
labels. Too brief and it cannot be used; too durable and it cannot be
reassigned:

\begin{equation}
\tau_{\mathrm{read}} \;\le\; \tau_{\mathrm{coh}} \;\le\; \tau_{\mathrm{update}}.\tag{6}
\end{equation}

For perceptual labelling in cortex, reading occurs within a percept, of
order 100 ms, and reconfiguration follows the theta and alpha cycle, of
order 100 to 250 ms, giving a window of roughly
\(\tau_{\mathrm{coh}} \in [0.05,\ 0.5]\) s.

This criterion cuts in the opposite direction from the usual worry. The
standard objection to exotic carriers is that coherence is too
\emph{fragile}; C5 adds that coherence can be too \emph{brief to be
useful even if achieved}, and, at the other end, too \emph{stable to be
reassigned}. The consequence is direct. A mechanism claiming millisecond
coherence has not thereby answered the objection, because milliseconds
remain two orders of magnitude below the window.

\hypertarget{c6.-writability}{%
\subsection{C6. Writability}\label{c6.-writability}}

A label is useless unless something sets it. The criterion requires a
physical mechanism that can shift the carrier across the usable range,
on the timescale of \(\tau_{\mathrm{update}}\), under the control of
whatever determines what the label ought to be.

This is the criterion most often absent altogether, and it is not a
formality. A carrier whose frequency is fixed by immutable molecular
properties cannot be a label at all, however high its \(Q\), because
every instance carries the same value. Writability is what distinguishes
a label from a constant, and it is the criterion on which the
neurochemical control of gamma frequency discussed in §1.1 bears most
directly.

\hypertarget{c7.-thermal-regime-preserved-or-driven}{%
\subsection{C7. Thermal regime: preserved or
driven}\label{c7.-thermal-regime-preserved-or-driven}}

A maintained collective state can be reached by two routes, and they are
alternatives rather than complements. A \emph{preserved} state is
protected by an energy gap large against \(kT\) and requires little
maintenance power. A \emph{driven and dissipative} state is continuously
re-established by pumping, as a laser above threshold, requiring no gap
but paying for it in C4.

A proposal must commit to one. The recurrent inconsistency is to claim
gap protection while the gap is a small fraction of \(kT\), and
simultaneously to assume no maintenance cost, thereby satisfying neither
branch. Where the gap is much smaller than \(kT\), C4 must be evaluated
in the driven regime.

The microwave proposal illustrates both halves of the difficulty. Its
gap argument requires the thermally vulnerable fraction to be
\(N_{\mathrm{vul}}/N \approx 10^{-3}\), which at 12 mM, where the mean
intermolecular spacing is 5.17 nm, corresponds to a vulnerable shell
exactly one molecule thick on a 30 \textmu{}m sphere. At three layers the margin
falls to 0.64 and at five it exceeds unity outright (Appendix B). Since
the boundary is a diffuse interface in liquid water rather than a solid
surface, confinement to one or two molecular layers is a strong
assumption, and it is neither stated nor justified.

There is also a category error worth isolating, because it recurs well
beyond this one model. The claim that a collective gap exceeds the
thermal energy admitted through a domain surface is an argument about
\emph{energy budget}, and it establishes thermodynamic stability.
Decoherence is a \emph{rate}. A system can sit in a deep collective
minimum and still lose phase coherence rapidly, because dephasing is
elastic: it requires only that which-path information leak, not that
energy be absorbed. The distinction is not ours. Reimers et al.~(2009)
reached it in the closely analogous Fröhlich setting, analysing the
Wu-Austin Hamiltonian (Wu and Austin, 1977; see also Bolterauer, 1999)
and reporting that rates of energy flow are surpassed in significance by
the rate of phase decoherence. Their simulations supply the relevant
number: in a fully formed strong condensate, coherence lifetime remains
on femtosecond timescales, shorter than a single vibrational period.
Collective ordering and phase coherence are separable, and only the
former is achieved.

Their three-way classification is the general form of C7 and we adopt
it: \emph{weak} condensates, in which effects on chemical kinetics are
possible; \emph{strong} condensates, in which a large amount of energy
is channelled into one vibrational mode, explicitly without coherence;
and \emph{coherent} condensates, in which that energy occupies a single
quantum state. A proposal should say which it claims, since the three
differ in what they can support and in what they cost. §6.1 returns to
where the microwave proposal falls.

\begin{center}\rule{0.5\linewidth}{0.5pt}\end{center}

\hypertarget{the-screen}{%
\section{The screen}\label{the-screen}}

The criteria are cheap enough to apply that a candidate can be screened
from two published quantities. Table 3 does this for the carriers
currently under discussion, using \(Q\) for C1 and the persistence
window for C5, the two criteria that discriminate. Gamma coherence times
follow Tan et al.~(2016).

\begin{longtable}[]{@{}lllllll@{}}
\caption{\textbf{Table 3.} The screen applied. \(Q\) computed from Eq.
(1); C5 evaluated against \(\tau_{\mathrm{coh}} \in [0.05, 0.5]\)
s.}\tabularnewline
\toprule
\begin{minipage}[b]{0.12\columnwidth}\raggedright
Carrier\strut
\end{minipage} & \begin{minipage}[b]{0.12\columnwidth}\raggedright
\(\nu\)\strut
\end{minipage} & \begin{minipage}[b]{0.12\columnwidth}\raggedright
\(\tau_{\mathrm{coh}}\)\strut
\end{minipage} & \begin{minipage}[b]{0.12\columnwidth}\raggedright
\(Q = M_{\max}\)\strut
\end{minipage} & \begin{minipage}[b]{0.12\columnwidth}\raggedright
C1\strut
\end{minipage} & \begin{minipage}[b]{0.12\columnwidth}\raggedright
C5\strut
\end{minipage} & \begin{minipage}[b]{0.12\columnwidth}\raggedright
note\strut
\end{minipage}\tabularnewline
\midrule
\endfirsthead
\toprule
\begin{minipage}[b]{0.12\columnwidth}\raggedright
Carrier\strut
\end{minipage} & \begin{minipage}[b]{0.12\columnwidth}\raggedright
\(\nu\)\strut
\end{minipage} & \begin{minipage}[b]{0.12\columnwidth}\raggedright
\(\tau_{\mathrm{coh}}\)\strut
\end{minipage} & \begin{minipage}[b]{0.12\columnwidth}\raggedright
\(Q = M_{\max}\)\strut
\end{minipage} & \begin{minipage}[b]{0.12\columnwidth}\raggedright
C1\strut
\end{minipage} & \begin{minipage}[b]{0.12\columnwidth}\raggedright
C5\strut
\end{minipage} & \begin{minipage}[b]{0.12\columnwidth}\raggedright
note\strut
\end{minipage}\tabularnewline
\midrule
\endhead
\begin{minipage}[t]{0.12\columnwidth}\raggedright
Cortical gamma\strut
\end{minipage} & \begin{minipage}[t]{0.12\columnwidth}\raggedright
40 Hz\strut
\end{minipage} & \begin{minipage}[t]{0.12\columnwidth}\raggedright
0.15 s\strut
\end{minipage} & \begin{minipage}[t]{0.12\columnwidth}\raggedright
38\strut
\end{minipage} & \begin{minipage}[t]{0.12\columnwidth}\raggedright
pass\strut
\end{minipage} & \begin{minipage}[t]{0.12\columnwidth}\raggedright
pass\strut
\end{minipage} & \begin{minipage}[t]{0.12\columnwidth}\raggedright
near centre of window\strut
\end{minipage}\tabularnewline
\begin{minipage}[t]{0.12\columnwidth}\raggedright
Cortical gamma, sustained\strut
\end{minipage} & \begin{minipage}[t]{0.12\columnwidth}\raggedright
40 Hz\strut
\end{minipage} & \begin{minipage}[t]{0.12\columnwidth}\raggedright
0.30 s\strut
\end{minipage} & \begin{minipage}[t]{0.12\columnwidth}\raggedright
75\strut
\end{minipage} & \begin{minipage}[t]{0.12\columnwidth}\raggedright
pass\strut
\end{minipage} & \begin{minipage}[t]{0.12\columnwidth}\raggedright
pass\strut
\end{minipage} & \begin{minipage}[t]{0.12\columnwidth}\raggedright
\strut
\end{minipage}\tabularnewline
\begin{minipage}[t]{0.12\columnwidth}\raggedright
Cortical alpha\strut
\end{minipage} & \begin{minipage}[t]{0.12\columnwidth}\raggedright
10 Hz\strut
\end{minipage} & \begin{minipage}[t]{0.12\columnwidth}\raggedright
0.5 s\strut
\end{minipage} & \begin{minipage}[t]{0.12\columnwidth}\raggedright
31\strut
\end{minipage} & \begin{minipage}[t]{0.12\columnwidth}\raggedright
pass\strut
\end{minipage} & \begin{minipage}[t]{0.12\columnwidth}\raggedright
pass\strut
\end{minipage} & \begin{minipage}[t]{0.12\columnwidth}\raggedright
at window edge\strut
\end{minipage}\tabularnewline
\begin{minipage}[t]{0.12\columnwidth}\raggedright
Hippocampal ripple\strut
\end{minipage} & \begin{minipage}[t]{0.12\columnwidth}\raggedright
180 Hz\strut
\end{minipage} & \begin{minipage}[t]{0.12\columnwidth}\raggedright
0.05 s\strut
\end{minipage} & \begin{minipage}[t]{0.12\columnwidth}\raggedright
57\strut
\end{minipage} & \begin{minipage}[t]{0.12\columnwidth}\raggedright
pass\strut
\end{minipage} & \begin{minipage}[t]{0.12\columnwidth}\raggedright
pass\strut
\end{minipage} & \begin{minipage}[t]{0.12\columnwidth}\raggedright
event-like\strut
\end{minipage}\tabularnewline
\begin{minipage}[t]{0.12\columnwidth}\raggedright
Cortical theta\strut
\end{minipage} & \begin{minipage}[t]{0.12\columnwidth}\raggedright
6 Hz\strut
\end{minipage} & \begin{minipage}[t]{0.12\columnwidth}\raggedright
1.0 s\strut
\end{minipage} & \begin{minipage}[t]{0.12\columnwidth}\raggedright
38\strut
\end{minipage} & \begin{minipage}[t]{0.12\columnwidth}\raggedright
pass\strut
\end{minipage} & \begin{minipage}[t]{0.12\columnwidth}\raggedright
fail\strut
\end{minipage} & \begin{minipage}[t]{0.12\columnwidth}\raggedright
too slow to rewrite\strut
\end{minipage}\tabularnewline
\begin{minipage}[t]{0.12\columnwidth}\raggedright
Calcium wave\strut
\end{minipage} & \begin{minipage}[t]{0.12\columnwidth}\raggedright
0.1 Hz\strut
\end{minipage} & \begin{minipage}[t]{0.12\columnwidth}\raggedright
100 s\strut
\end{minipage} & \begin{minipage}[t]{0.12\columnwidth}\raggedright
63\strut
\end{minipage} & \begin{minipage}[t]{0.12\columnwidth}\raggedright
pass\strut
\end{minipage} & \begin{minipage}[t]{0.12\columnwidth}\raggedright
fail\strut
\end{minipage} & \begin{minipage}[t]{0.12\columnwidth}\raggedright
far too slow\strut
\end{minipage}\tabularnewline
\begin{minipage}[t]{0.12\columnwidth}\raggedright
Protein vibrational mode\strut
\end{minipage} & \begin{minipage}[t]{0.12\columnwidth}\raggedright
10 THz\strut
\end{minipage} & \begin{minipage}[t]{0.12\columnwidth}\raggedright
1 ps\strut
\end{minipage} & \begin{minipage}[t]{0.12\columnwidth}\raggedright
63\strut
\end{minipage} & \begin{minipage}[t]{0.12\columnwidth}\raggedright
pass\strut
\end{minipage} & \begin{minipage}[t]{0.12\columnwidth}\raggedright
fail\strut
\end{minipage} & \begin{minipage}[t]{0.12\columnwidth}\raggedright
\(5\times10^{10}\) too brief\strut
\end{minipage}\tabularnewline
\begin{minipage}[t]{0.12\columnwidth}\raggedright
Microtubule terahertz mode\strut
\end{minipage} & \begin{minipage}[t]{0.12\columnwidth}\raggedright
10 THz\strut
\end{minipage} & \begin{minipage}[t]{0.12\columnwidth}\raggedright
0.1 ps\strut
\end{minipage} & \begin{minipage}[t]{0.12\columnwidth}\raggedright
6.3\strut
\end{minipage} & \begin{minipage}[t]{0.12\columnwidth}\raggedright
marginal\strut
\end{minipage} & \begin{minipage}[t]{0.12\columnwidth}\raggedright
fail\strut
\end{minipage} & \begin{minipage}[t]{0.12\columnwidth}\raggedright
published femtosecond coherence\strut
\end{minipage}\tabularnewline
\begin{minipage}[t]{0.12\columnwidth}\raggedright
\textbf{Microwave carrier, generous proxy}\strut
\end{minipage} & \begin{minipage}[t]{0.12\columnwidth}\raggedright
\textbf{30 GHz}\strut
\end{minipage} & \begin{minipage}[t]{0.12\columnwidth}\raggedright
\textbf{1 ps}\strut
\end{minipage} & \begin{minipage}[t]{0.12\columnwidth}\raggedright
\textbf{0.19}\strut
\end{minipage} & \begin{minipage}[t]{0.12\columnwidth}\raggedright
\textbf{fail}\strut
\end{minipage} & \begin{minipage}[t]{0.12\columnwidth}\raggedright
\textbf{fail}\strut
\end{minipage} & \begin{minipage}[t]{0.12\columnwidth}\raggedright
not an oscillator\strut
\end{minipage}\tabularnewline
\begin{minipage}[t]{0.12\columnwidth}\raggedright
Microwave carrier, continuum estimate\strut
\end{minipage} & \begin{minipage}[t]{0.12\columnwidth}\raggedright
30 GHz\strut
\end{minipage} & \begin{minipage}[t]{0.12\columnwidth}\raggedright
27 fs\strut
\end{minipage} & \begin{minipage}[t]{0.12\columnwidth}\raggedright
0.005\strut
\end{minipage} & \begin{minipage}[t]{0.12\columnwidth}\raggedright
fail\strut
\end{minipage} & \begin{minipage}[t]{0.12\columnwidth}\raggedright
fail\strut
\end{minipage} & \begin{minipage}[t]{0.12\columnwidth}\raggedright
appropriate value at 260 cm\(^{-1}\)\strut
\end{minipage}\tabularnewline
\begin{minipage}[t]{0.12\columnwidth}\raggedright
Microwave carrier, as its model requires\strut
\end{minipage} & \begin{minipage}[t]{0.12\columnwidth}\raggedright
30 GHz\strut
\end{minipage} & \begin{minipage}[t]{0.12\columnwidth}\raggedright
1 ms\strut
\end{minipage} & \begin{minipage}[t]{0.12\columnwidth}\raggedright
\(1.9\times10^{8}\)\strut
\end{minipage} & \begin{minipage}[t]{0.12\columnwidth}\raggedright
pass\strut
\end{minipage} & \begin{minipage}[t]{0.12\columnwidth}\raggedright
fail\strut
\end{minipage} & \begin{minipage}[t]{0.12\columnwidth}\raggedright
still \(50\times\) too brief\strut
\end{minipage}\tabularnewline
\bottomrule
\end{longtable}

Four observations follow.

\emph{Only the low-frequency neural oscillations pass both C1 and C5.}
This is the framework's main structural finding, and it is not what the
usual framing of these debates would lead one to expect. High-frequency
molecular carriers satisfy C1 comfortably, since a terahertz mode with
picosecond coherence still executes some sixty cycles, and then fail C5
by ten or more orders of magnitude. They are distinguishable without
being useful. The constraint that eliminates them is not fragility but
brevity, and no improvement in isolation or shielding addresses it.

\emph{Granting the microwave proposal everything it asks for does not
save it.} The final row takes the millisecond coherence time the model
requires, rather than any measured value, and the carrier then passes C1
by eight orders of magnitude while still failing C5 by a factor of
fifty. §2 rejected the proposal on C1, C3 and C4; the screen adds that
even a complete victory on those three would leave it short on
persistence. Four independent failures suggests the difficulty lies with
the choice of carrier rather than with any particular parameter.

\emph{The terahertz coherence times in Table 3 are upper bounds granted
to the proposals, not measurements supporting them.} They rest on the
same carboxylate-stretch substitution discussed in §2.3, and the
appropriate values at the relevant mode frequencies are shorter by one
to two orders of magnitude. Every terahertz entry is therefore more
favourable than the physics warrants.

\emph{Overdelivery is a warning sign.} A carrier that must be pushed to
extreme parameters to satisfy C1, and that then supplies \(10^8\)
channels where tens are needed, is probably the wrong mechanism.
Biological solutions tend to sit close to their constraints rather than
many orders above them, and a mechanism that is either impossible or
wildly excessive, with nothing in between, is usually being asked to do
a job that something else already does.

\hypertarget{why-biology-mostly-does-not-use-frequency-labels}{%
\subsection{Why biology mostly does not use frequency
labels}\label{why-biology-mostly-does-not-use-frequency-labels}}

That last point generalises, and the bound explains a pattern that is
otherwise easy to overlook, set out in Table 4.

\begin{longtable}[]{@{}lll@{}}
\caption{\textbf{Table 4.} Approximate alphabet size of biological
coding schemes, and what limits each.}\tabularnewline
\toprule
Scheme & Approximate alphabet & Bounded by\tabularnewline
\midrule
\endfirsthead
\toprule
Scheme & Approximate alphabet & Bounded by\tabularnewline
\midrule
\endhead
Chemical identity & \(10^3\) to \(10^6\) & molecular shape and receptor
specificity\tabularnewline
Spatial address by wiring & \(10^4\) to \(10^{11}\) & which axon
contacts which target\tabularnewline
Spike timing & \(10^2\) to \(10^3\) & jitter relative to the integration
window\tabularnewline
\textbf{Frequency multiplex} & \textbf{10 to 100} & \textbf{\(Q\)}, Eq.
(1)\tabularnewline
Rate code & 10 to 100 & count noise in the integration
window\tabularnewline
\bottomrule
\end{longtable}

Biology uses chemistry and wiring where it needs a large alphabet, and
reserves frequency and rate for small ones. On the account given here
this is not an accident of evolutionary history but a consequence of C1:
wet tissue at 310 K cannot host high-\(Q\) resonators, so frequency is
intrinsically a low-capacity channel, and any organism needing \(10^6\)
distinguishable labels must obtain them elsewhere.

The consequence for proposals is a question they should be expected to
answer. A mechanism that asks frequency to supply a large alphabet is
asking it to do a job that biology assigns to chemistry and wiring, and
should explain why the usual solution was unavailable in the case at
hand. We are not aware of a proposal in this literature that addresses
the question.

\begin{center}\rule{0.5\linewidth}{0.5pt}\end{center}

\hypertarget{what-a-viable-proposal-would-look-like}{%
\section{What a viable proposal would look
like}\label{what-a-viable-proposal-would-look-like}}

The screen is easier to state positively, as a specification. A
candidate carrier should arrive with seven things:

\begin{enumerate}
\def\labelenumi{\arabic{enumi}.}
\tightlist
\item
  A carrier frequency and a measured coherence time, from which \(Q\),
  and hence the maximum channel count, follows immediately. If
  \(Q < 1\), stop.
\item
  A required channel count, derived from the function claimed and
  checked against \(Q\). If the two differ by orders of magnitude in
  either direction, explain why.
\item
  A coupling mechanism with a stated range, and, if radiative, an
  emitting structure of size comparable to the wavelength.
\item
  A power budget: energy in the mode divided by coherence time, against
  the metabolic supply of the tissue volume involved.
\item
  A position in the persistence window: read time, hold time, rewrite
  time.
\item
  A writing mechanism: what sets the label, across what range, how fast.
\item
  A declared thermal regime, gap-protected or driven, with the
  corresponding power consequence accepted.
\end{enumerate}

This is not a hypothetical procedure. It is what §2 carried out, and the
specification is offered in the expectation that applying it takes an
afternoon rather than a research programme. A screen whose application
requires simulation is a screen that will not be applied, and the
arithmetic here is deliberately elementary for that reason: every item
can be evaluated on paper from two or three published quantities. A
reference implementation of the seven criteria, with scripts that
regenerate every table here, is archived as Kopel (2026b).

\hypertarget{a-design-rule}{%
\subsection{A design rule}\label{a-design-rule}}

Criteria C1 and C5 combine. Since C5 caps the coherence time from above
at \(\tau_{\mathrm{update}}\), and C1 caps the channel count at
\(2\pi\nu\tau_{\mathrm{coh}}\),

\begin{equation}
M \;\lesssim\; 2\pi\nu\,\tau_{\mathrm{update}}.\tag{7}
\end{equation}

Channel count therefore grows with carrier frequency, which is the
intuition that motivates high-frequency proposals in the first place,
but only until coherence fails the \emph{read} requirement at the other
end of the window. Eq. (7) is free of any assumption about how long
biological oscillators actually ring.

The complementary ceiling is not. If oscillators sustain approximately
\(N_{\mathrm{cyc}}\) cycles of phase coherence, then the highest usable
carrier is

\begin{equation}
\nu_{\max} \;\approx\; \frac{N_{\mathrm{cyc}}}{\tau_{\mathrm{read}}},\tag{8}
\end{equation}

which for \(N_{\mathrm{cyc}} \approx 10\) and
\(\tau_{\mathrm{read}} \approx 0.1\) s gives \(\nu_{\max} \approx 100\)
Hz. Cortical gamma sits just below that ceiling. On this reading gamma
is not an arbitrary band but approximately the highest frequency at
which cortical tissue can hold a label for the duration of a percept,
which would be an explanation rather than a description.

We flag the status of that inference carefully, because it is the most
attractive claim in the paper and the least secure. It depends entirely
on \(N_{\mathrm{cyc}}\), a quantity estimated rather than measured
across most of the range it would need to hold over, and §6.3 sets out
why we treat it as unresolved. Eq. (7) stands without it; Eq. (8), and
the account of gamma that follows from it, do not.

\begin{center}\rule{0.5\linewidth}{0.5pt}\end{center}

\hypertarget{discussion}{%
\section{Discussion}\label{discussion}}

\hypertarget{what-the-analysis-of-2-establishes-and-what-it-does-not}{%
\subsection{What the analysis of §2 establishes, and what it does
not}\label{what-the-analysis-of-2-establishes-and-what-it-does-not}}

The conclusion is narrower than a dismissal, and the distinction
matters. Following the classification adopted as C7, our analysis places
the glutamate domain in the \emph{weak} regime: a rate-equation
condensate, excluded from the coherent regime by five to nine orders of
magnitude on power alone, and marginal at best in the strong regime.
This is the same reclassification Reimers et al.~(2009) applied to the
coherent-QED programme of Del Giudice et al.~(2005) on which the model
is built.

The weak reading is not nothing. A condensate that funnels energy into a
frequency-selective collective mode produces real effects on chemical
kinetics and constitutes a genuine mode-selection mechanism. What it
does not produce is macroscopic quantum coherence, a millisecond-lived
collective state, or a high-\(Q\) microwave label. Only the last of
these is required by the argument in §2, and only the last is refuted.

The decoherence calculation identified as necessary future work by
Keppler (2023) has not been supplied in the papers that followed
(Keppler, 2024, 2025, 2026), and therefore remains worth doing. The most
recent of those states that the coupling mechanism ``still needs to be
empirically substantiated'' (Keppler, 2026), which places the
disagreement here more narrowly than it may appear: we differ over
whether the labelling function survives, not over whether the mechanism
is established. It is a well-posed problem in open-system dynamics and
it would close the programme's central gap in either direction. It is
simply no longer on the critical path for the \emph{labelling} question,
which §2 settles independently of its outcome.

\hypertarget{a-structural-insight-worth-preserving}{%
\subsection{A structural insight worth
preserving}\label{a-structural-insight-worth-preserving}}

Before the criticism is allowed to stand as the whole verdict, one
feature of the model deserves to be extracted from it, because it
survives the failure of the physics that motivated it.

The model contains two independently regulated concentration parameters
doing two different jobs. Its ignition criterion is evaluated at the
\emph{vesicular} glutamate concentration, while the stationary-state
frequency is evaluated at the \emph{tissue} concentration. These are
distinct variables governed by distinct molecular machinery: vesicular
loading by neuronal transporters, bulk pool concentration by astrocytic
uptake and the glutamate-glutamine cycle. A perturbation confined to
astrocytic clearance therefore shifts the carrier frequency while
leaving the capacity to generate a collective mode untouched, and vice
versa. The model thus contains a gain axis and a tuning axis that are
quasi-orthogonal by anatomy, and that fail in categorically different
ways: loss of gain removes the mode, whereas loss of tuning leaves it
intact but \emph{mislabelled}.

The source treats the stationary state purely as a consequence of
ignition and never separates the two. The separation is
substrate-independent, requires only \emph{some} frequency-selective
collective mode, and does not require that mode to be quantum,
microwave, or long-lived. It is developed elsewhere with a carrier that
passes the screen of §4.

\hypertarget{an-open-question-does-q-depend-on-frequency}{%
\subsection{\texorpdfstring{An open question: does \(Q\) depend on
frequency?}{An open question: does Q depend on frequency?}}\label{an-open-question-does-q-depend-on-frequency}}

Assembling Table 3, one notices that every physiological carrier lands
at \(Q \approx 6\) to 75 across fourteen decades of carrier frequency.
If real, the regularity would have a clean consequence: any
frequency-multiplexed biological code would be capped near 10 to 100
labels regardless of carrier, because coherence would degrade in
proportion to frequency and the ratio would stay fixed. This is the
\(N_{\mathrm{cyc}}\) on which Eq. (8) depends.

We do not present it as a result, because the procedure that produced it
was partly circular. Several coherence times were estimated as a few to
a few tens of cycles of the carrier itself, and given that input,
\(Q = 2\pi \times (\text{cycles})\) follows tautologically. The apparent
scale-invariance is substantially an artefact of how the inputs were
chosen.

The test is nonetheless straightforward and already runnable: assemble
\emph{fitted damping rates}, rather than cycle-count estimates or the
widths of time-averaged spectra, across as many decades as possible, and
plot \(Q\) against \(\nu\). Four published datasets already supply such
measurements. Spyropoulos et al.~(2022) fit macaque V1 gamma as a
noise-driven damped harmonic oscillator, from which a damping
coefficient yields \(Q\) directly. Pesnot Lerousseau et al.~(2021) fit
human intracranial and surface recordings with a
damped-harmonic-oscillator model and classify responses into dynamical
classes with distinct damping and eigenfrequencies. Blanco-Duque et
al.~(2024) define an explicit oscillatory-quality metric for sleep
spindles from autoregressive fits. Casanova et al.~(2021) use the
ringing decay of gamma envelopes as a damping measure.

Two of these argue against the conjecture. Pesnot Lerousseau et
al.~(2021) report persistent high-gamma activity outlasting stimulation
throughout cortex while a 2.5 Hz stream produced no persistent activity
in any band, which would mean \(Q\) rises with frequency. And
Blanco-Duque et al.~(2024) find that oscillatory quality varies
substantially with lamina, region and sleep-wake history: a quantity
that is not constant within one rhythm is unlikely to be constant across
fourteen decades.

The conjecture is therefore not merely untested; it is testable now, and
the preliminary indications are that it is probably false as stated.
What survives either way is Eq. (1), which is a derivation rather than
an observation, together with the practical consequence that \(Q\) must
be \emph{measured} for each candidate carrier rather than inferred from
burst duration. If \(Q\) does rise with frequency, the interesting
question changes from why \(Q\) is fixed to what sets its frequency
dependence, and whether any biological carrier reaches a \(Q\) high
enough to matter.

\hypertarget{what-would-overturn-this-analysis}{%
\subsection{What would overturn this
analysis}\label{what-would-overturn-this-analysis}}

Stated so that the argument can be attacked at its weakest points.

\begin{itemize}
\tightlist
\item
  Vibrational \(T_2\) in the glutamate-water matrix exceeds
  bulk-solution values by \(\gtrsim 10^3\): §2.2 relaxes and a microwave
  label becomes possible. An extraordinary claim, but a testable one.
\item
  A rigorous decoherence calculation yields \(\tau \gtrsim 1\) \textmu{}s: §2.2
  and C7 both relax substantially, and Eq. (3) then permits some
  \(10^{7}\) quanta rather than fewer than twelve.
\item
  The vulnerable shell is justified at one molecular layer on physical
  grounds: the C7 objection is resolved.
\item
  Broadband terahertz spectroscopy confirms a strong hydrated-glutamate
  line at 7.8 THz: the model's frequency selection is placed on measured
  footing (see Appendix C).
\item
  A biological oscillator is measured at \(Q \gg 100\): §4.1's account
  of why biology reserves frequency for small alphabets weakens.
\item
  Measured \(Q\) proves flat against \(\nu\) across many decades: the
  conjecture of §6.3 is upheld and Eq. (8) becomes a result.
\end{itemize}

\begin{center}\rule{0.5\linewidth}{0.5pt}\end{center}

\hypertarget{limits-of-the-framework}{%
\section{Limits of the framework}\label{limits-of-the-framework}}

\emph{It screens frequency-multiplexed codes only.} Carriers that encode
in amplitude, in phase relative to a reference, in chemical identity or
in spatial address require different bounds. Eq. (1) does not apply to
them, and a proposal rejected under C1 might be rescued by reformulating
what carries the information rather than by defending the carrier.

\emph{The criteria are necessary, not sufficient.} Passing all seven
establishes that a mechanism \emph{could} serve as a label, not that it
does. Cortical gamma passes, which is consistent with a labelling role
and does not demonstrate one. Ray and Maunsell (2010) show that gamma
frequency tracks stimulus contrast and differs measurably between
assemblies a few hundred micrometres apart, and conclude that gamma is a
poor candidate for binding or communication in V1. Nothing here touches
that argument: it concerns whether a carrier that \emph{could} label in
fact does, which is a question the screen is not designed to answer.

\emph{The persistence window is task-specific.} We computed it for
perceptual labelling. Working memory, motor control and consolidation
impose different read and update times, and C5 must be recomputed for
each. A carrier failing C5 for one task may satisfy it for another.

\emph{Coherence time carries most of the weight and is often the least
secure number in the calculation.} For the majority of candidates in
Table 3 it is estimated rather than measured, which is the same weakness
§6.3 identifies in the conjecture. Where a proposal turns on \(Q\), the
coherence time should be measured directly rather than inferred from
burst duration.

\emph{The depth of scrutiny is uneven.} We audited one proposal in
detail and screened the remainder at table resolution, taking published
coherence estimates at face value. The microtubule entries in particular
deserve the treatment §2 gives the microwave carrier before their
rejection is regarded as settled.

\emph{We have not addressed noise correlation.} C2 assumes independent
noise across readers. Correlated noise would change the effective
channel count in ways this framework does not capture.

\begin{center}\rule{0.5\linewidth}{0.5pt}\end{center}

\hypertarget{appendix-a.-reproduction-of-the-source-models-arithmetic}{%
\section*{Appendix A. Reproduction of the source model's
arithmetic}\label{appendix-a.-reproduction-of-the-source-models-arithmetic}}
\addcontentsline{toc}{section}{Appendix A. Reproduction of the source
model's arithmetic}

All quantities below were recomputed from the defining relations using
\(c = 2.99792458\times10^{8}\) m s\(^{-1}\),
\(N_A = 6.02214076\times10^{23}\) mol\(^{-1}\) and
\(k_B = 8.617333\times10^{-5}\) eV K\(^{-1}\), at \(T = 310\) K.
Glutamate concentrations are taken from the source model: 300 mM
vesicular, 12 mM tissue. The coherence domain is treated as a sphere of
diameter \(d = 30\) \textmu{}m. Table A.1 sets the reported values against the
recomputed ones.

\begin{longtable}[]{@{}llll@{}}
\caption{\textbf{Table A.1.} Internal arithmetic of the source model, as
reported and as recomputed.}\tabularnewline
\toprule
\begin{minipage}[b]{0.22\columnwidth}\raggedright
Quantity\strut
\end{minipage} & \begin{minipage}[b]{0.22\columnwidth}\raggedright
Reported\strut
\end{minipage} & \begin{minipage}[b]{0.22\columnwidth}\raggedright
Recomputed\strut
\end{minipage} & \begin{minipage}[b]{0.22\columnwidth}\raggedright
Status\strut
\end{minipage}\tabularnewline
\midrule
\endfirsthead
\toprule
\begin{minipage}[b]{0.22\columnwidth}\raggedright
Quantity\strut
\end{minipage} & \begin{minipage}[b]{0.22\columnwidth}\raggedright
Reported\strut
\end{minipage} & \begin{minipage}[b]{0.22\columnwidth}\raggedright
Recomputed\strut
\end{minipage} & \begin{minipage}[b]{0.22\columnwidth}\raggedright
Status\strut
\end{minipage}\tabularnewline
\midrule
\endhead
\begin{minipage}[t]{0.22\columnwidth}\raggedright
Coupling scaling with concentration\strut
\end{minipage} & \begin{minipage}[t]{0.22\columnwidth}\raggedright
``one fifth'' at tissue concentration\strut
\end{minipage} & \begin{minipage}[t]{0.22\columnwidth}\raggedright
\(\sqrt{12/300} = 0.2000\)\strut
\end{minipage} & \begin{minipage}[t]{0.22\columnwidth}\raggedright
confirms \(g \propto \sqrt{n}\)\strut
\end{minipage}\tabularnewline
\begin{minipage}[t]{0.22\columnwidth}\raggedright
\(\lambda\) at \(\nu_0 = 7.8\) THz\strut
\end{minipage} & \begin{minipage}[t]{0.22\columnwidth}\raggedright
38.4 \textmu{}m\strut
\end{minipage} & \begin{minipage}[t]{0.22\columnwidth}\raggedright
38.435 \textmu{}m\strut
\end{minipage} & \begin{minipage}[t]{0.22\columnwidth}\raggedright
agrees\strut
\end{minipage}\tabularnewline
\begin{minipage}[t]{0.22\columnwidth}\raggedright
Domain diameter as \(\tfrac{\pi}{4}\lambda(\nu_0)\)\strut
\end{minipage} & \begin{minipage}[t]{0.22\columnwidth}\raggedright
30 \textmu{}m\strut
\end{minipage} & \begin{minipage}[t]{0.22\columnwidth}\raggedright
30.187 \textmu{}m\strut
\end{minipage} & \begin{minipage}[t]{0.22\columnwidth}\raggedright
agrees to 0.6\%\strut
\end{minipage}\tabularnewline
\begin{minipage}[t]{0.22\columnwidth}\raggedright
\(d/\lambda(\nu_0)\) against \(\pi/4\)\strut
\end{minipage} & \begin{minipage}[t]{0.22\columnwidth}\raggedright
resonant\strut
\end{minipage} & \begin{minipage}[t]{0.22\columnwidth}\raggedright
0.7805 vs 0.7854\strut
\end{minipage} & \begin{minipage}[t]{0.22\columnwidth}\raggedright
agrees to 0.6\%\strut
\end{minipage}\tabularnewline
\begin{minipage}[t]{0.22\columnwidth}\raggedright
Number density at 12 mM\strut
\end{minipage} & \begin{minipage}[t]{0.22\columnwidth}\raggedright
not stated\strut
\end{minipage} & \begin{minipage}[t]{0.22\columnwidth}\raggedright
\(7.227\times10^{24}\) m\(^{-3}\)\strut
\end{minipage} & \begin{minipage}[t]{0.22\columnwidth}\raggedright
see Appendix B\strut
\end{minipage}\tabularnewline
\begin{minipage}[t]{0.22\columnwidth}\raggedright
Molecules per domain\strut
\end{minipage} & \begin{minipage}[t]{0.22\columnwidth}\raggedright
\(\sim10^{11}\)\strut
\end{minipage} & \begin{minipage}[t]{0.22\columnwidth}\raggedright
\(1.022\times10^{11}\)\strut
\end{minipage} & \begin{minipage}[t]{0.22\columnwidth}\raggedright
agrees\strut
\end{minipage}\tabularnewline
\begin{minipage}[t]{0.22\columnwidth}\raggedright
\(k_BT\) at 310 K\strut
\end{minipage} & \begin{minipage}[t]{0.22\columnwidth}\raggedright
26 meV\strut
\end{minipage} & \begin{minipage}[t]{0.22\columnwidth}\raggedright
26.71 meV\strut
\end{minipage} & \begin{minipage}[t]{0.22\columnwidth}\raggedright
agrees\strut
\end{minipage}\tabularnewline
\begin{minipage}[t]{0.22\columnwidth}\raggedright
Per-molecule gap against \(k_BT\)\strut
\end{minipage} & \begin{minipage}[t]{0.22\columnwidth}\raggedright
gap protects\strut
\end{minipage} & \begin{minipage}[t]{0.22\columnwidth}\raggedright
\(0.130\) meV \(= k_BT/205\)\strut
\end{minipage} & \begin{minipage}[t]{0.22\columnwidth}\raggedright
reproduced; see note\strut
\end{minipage}\tabularnewline
\bottomrule
\end{longtable}

Two remarks on the last row. The ratio is reproduced exactly, but its
interpretation is not the source model's: a Boltzmann factor of
\(\exp(-0.130/26.71) = 0.995\) leaves no single-molecule thermal
protection whatever, so the entire protection claim rests on
collectivity, which is the subject of Appendix B and of §3.6.

We record separately that the agreement on domain diameter is closer
than the source model claims. Our recomputation gives 0.6\% rather than
the larger figure quoted, which strengthens rather than weakens the
model's internal consistency, and we note it because §2.1 rests on that
consistency being real.

\begin{longtable}[]{@{}lllll@{}}
\caption{\textbf{Table A.2.} The same domain evaluated at both
frequencies the model assigns it, with \(d = 30\) \textmu{}m throughout. This is
the basis of the geometric argument in §2.4.}\tabularnewline
\toprule
\begin{minipage}[b]{0.17\columnwidth}\raggedright
\strut
\end{minipage} & \begin{minipage}[b]{0.17\columnwidth}\raggedright
wavelength\strut
\end{minipage} & \begin{minipage}[b]{0.17\columnwidth}\raggedright
\(d/\lambda\)\strut
\end{minipage} & \begin{minipage}[b]{0.17\columnwidth}\raggedright
\(ka = \pi d/\lambda\)\strut
\end{minipage} & \begin{minipage}[b]{0.17\columnwidth}\raggedright
\((ka)^2\)\strut
\end{minipage}\tabularnewline
\midrule
\endfirsthead
\toprule
\begin{minipage}[b]{0.17\columnwidth}\raggedright
\strut
\end{minipage} & \begin{minipage}[b]{0.17\columnwidth}\raggedright
wavelength\strut
\end{minipage} & \begin{minipage}[b]{0.17\columnwidth}\raggedright
\(d/\lambda\)\strut
\end{minipage} & \begin{minipage}[b]{0.17\columnwidth}\raggedright
\(ka = \pi d/\lambda\)\strut
\end{minipage} & \begin{minipage}[b]{0.17\columnwidth}\raggedright
\((ka)^2\)\strut
\end{minipage}\tabularnewline
\midrule
\endhead
\begin{minipage}[t]{0.17\columnwidth}\raggedright
\(\nu_0 = 7.8\) THz\strut
\end{minipage} & \begin{minipage}[t]{0.17\columnwidth}\raggedright
38.44 \textmu{}m\strut
\end{minipage} & \begin{minipage}[t]{0.17\columnwidth}\raggedright
0.7805 (\(\approx \pi/4\), resonant)\strut
\end{minipage} & \begin{minipage}[t]{0.17\columnwidth}\raggedright
2.45\strut
\end{minipage} & \begin{minipage}[t]{0.17\columnwidth}\raggedright
6.0\strut
\end{minipage}\tabularnewline
\begin{minipage}[t]{0.17\columnwidth}\raggedright
\(\nu_{\mathrm{coh}} = 30\) GHz\strut
\end{minipage} & \begin{minipage}[t]{0.17\columnwidth}\raggedright
9.99 mm\strut
\end{minipage} & \begin{minipage}[t]{0.17\columnwidth}\raggedright
0.00300, that is \(\lambda/333\)\strut
\end{minipage} & \begin{minipage}[t]{0.17\columnwidth}\raggedright
0.0094\strut
\end{minipage} & \begin{minipage}[t]{0.17\columnwidth}\raggedright
\(8.9\times10^{-5}\)\strut
\end{minipage}\tabularnewline
\bottomrule
\end{longtable}

The structure is defined to be resonant at \(\nu_0\) and is \(1/333\) of
a wavelength across at \(\nu_{\mathrm{coh}}\). The final column is given
only to support the scope limit stated in §2.4: the far-field radiation
efficiency is negligible at 30 GHz, but this bears on the availability
of a cavity and not on near-field action within the column.

\begin{center}\rule{0.5\linewidth}{0.5pt}\end{center}

\hypertarget{appendix-b.-the-vulnerable-shell-calculation}{%
\section*{Appendix B. The vulnerable-shell
calculation}\label{appendix-b.-the-vulnerable-shell-calculation}}
\addcontentsline{toc}{section}{Appendix B. The vulnerable-shell
calculation}

The source model's thermal-protection argument requires that the
thermally vulnerable fraction of the domain satisfy
\(N_{\mathrm{vul}}E_{\mathrm{th}} < \Delta E_{\mathrm{gap}}\), with
\(N_{\mathrm{vul}}/N \approx 10^{-3}\). The question is what shell
thickness that fraction corresponds to.

At 12 mM the number density is \(n = 7.227\times10^{24}\) m\(^{-3}\),
giving a mean intermolecular spacing of \(n^{-1/3} = 5.17\) nm. For a
sphere of radius \(r = 15\) \textmu{}m, surface area
\(A = 4\pi r^2 = 2.827\times10^{-9}\) m\(^2\), a shell of thickness
\(\delta\) contains \(N_{\mathrm{vul}} = nA\delta\) molecules. With
\(E_{\mathrm{th}} = k_BT = 26.71\) meV and
\(\Delta E_{\mathrm{gap}} = N \times 0.130\) meV \(= 1.328\times10^{7}\)
eV, the margin is as given in Table B.1:

\begin{longtable}[]{@{}lllll@{}}
\caption{\textbf{Table B.1.} Thermal margin as a function of shell
thickness, in units of the mean intermolecular spacing.}\tabularnewline
\toprule
\begin{minipage}[b]{0.17\columnwidth}\raggedright
Shell\strut
\end{minipage} & \begin{minipage}[b]{0.17\columnwidth}\raggedright
Thickness\strut
\end{minipage} & \begin{minipage}[b]{0.17\columnwidth}\raggedright
\(N_{\mathrm{vul}}\)\strut
\end{minipage} & \begin{minipage}[b]{0.17\columnwidth}\raggedright
\(N_{\mathrm{vul}}k_BT / \Delta E_{\mathrm{gap}}\)\strut
\end{minipage} & \begin{minipage}[b]{0.17\columnwidth}\raggedright
Verdict\strut
\end{minipage}\tabularnewline
\midrule
\endfirsthead
\toprule
\begin{minipage}[b]{0.17\columnwidth}\raggedright
Shell\strut
\end{minipage} & \begin{minipage}[b]{0.17\columnwidth}\raggedright
Thickness\strut
\end{minipage} & \begin{minipage}[b]{0.17\columnwidth}\raggedright
\(N_{\mathrm{vul}}\)\strut
\end{minipage} & \begin{minipage}[b]{0.17\columnwidth}\raggedright
\(N_{\mathrm{vul}}k_BT / \Delta E_{\mathrm{gap}}\)\strut
\end{minipage} & \begin{minipage}[b]{0.17\columnwidth}\raggedright
Verdict\strut
\end{minipage}\tabularnewline
\midrule
\endhead
\begin{minipage}[t]{0.17\columnwidth}\raggedright
1 layer\strut
\end{minipage} & \begin{minipage}[t]{0.17\columnwidth}\raggedright
5.2 nm\strut
\end{minipage} & \begin{minipage}[t]{0.17\columnwidth}\raggedright
\(1.06\times10^{8}\)\strut
\end{minipage} & \begin{minipage}[t]{0.17\columnwidth}\raggedright
0.21\strut
\end{minipage} & \begin{minipage}[t]{0.17\columnwidth}\raggedright
holds\strut
\end{minipage}\tabularnewline
\begin{minipage}[t]{0.17\columnwidth}\raggedright
2 layers\strut
\end{minipage} & \begin{minipage}[t]{0.17\columnwidth}\raggedright
10.3 nm\strut
\end{minipage} & \begin{minipage}[t]{0.17\columnwidth}\raggedright
\(2.11\times10^{8}\)\strut
\end{minipage} & \begin{minipage}[t]{0.17\columnwidth}\raggedright
0.43\strut
\end{minipage} & \begin{minipage}[t]{0.17\columnwidth}\raggedright
holds\strut
\end{minipage}\tabularnewline
\begin{minipage}[t]{0.17\columnwidth}\raggedright
3 layers\strut
\end{minipage} & \begin{minipage}[t]{0.17\columnwidth}\raggedright
15.5 nm\strut
\end{minipage} & \begin{minipage}[t]{0.17\columnwidth}\raggedright
\(3.17\times10^{8}\)\strut
\end{minipage} & \begin{minipage}[t]{0.17\columnwidth}\raggedright
0.64\strut
\end{minipage} & \begin{minipage}[t]{0.17\columnwidth}\raggedright
marginal\strut
\end{minipage}\tabularnewline
\begin{minipage}[t]{0.17\columnwidth}\raggedright
5 layers\strut
\end{minipage} & \begin{minipage}[t]{0.17\columnwidth}\raggedright
25.9 nm\strut
\end{minipage} & \begin{minipage}[t]{0.17\columnwidth}\raggedright
\(5.28\times10^{8}\)\strut
\end{minipage} & \begin{minipage}[t]{0.17\columnwidth}\raggedright
1.06\strut
\end{minipage} & \begin{minipage}[t]{0.17\columnwidth}\raggedright
fails\strut
\end{minipage}\tabularnewline
\bottomrule
\end{longtable}

The margin is linear in shell thickness, and the model's assumed
fraction corresponds to a shell exactly one molecule thick. The
protection argument is therefore sound only if thermal exchange is
confined to one or two molecular layers at the domain boundary. Since
that boundary is a diffuse interface in liquid water rather than a solid
surface, this is a strong assumption. It is not stated as one in the
source, and we could find no justification for it. It should be
supplied, or the argument weakened.

\begin{center}\rule{0.5\linewidth}{0.5pt}\end{center}

\hypertarget{appendix-c.-the-selection-of-7.8-thz}{%
\section*{Appendix C. The selection of 7.8
THz}\label{appendix-c.-the-selection-of-7.8-thz}}
\addcontentsline{toc}{section}{Appendix C. The selection of 7.8 THz}

Much of the source model's numerology descends from a single reference
frequency: the domain diameter is \(\tfrac{\pi}{4}\lambda(\nu_0)\), and
the microwave carrier is a renormalisation of \(\nu_0\). The provenance
of that frequency is therefore worth recording, and it is weaker than
the use made of it.

Three points, in ascending order of seriousness.

First, published absorption data for glutamate do not extend above
approximately 5 THz, so 7.8 THz lies outside the measured range. It is
not a measured line for this molecule.

Second, the underlying vibrational parameter was estimated from spectra
of \emph{dry} GABA rather than hydrated glutamate. This is a
substitution of both molecule and phase. The phase substitution is the
more consequential of the two, because the model's mechanism turns
specifically on hydration-enhanced dipole moments: hydration is what
lifts the coupling constant into the critical regime. Data from the dry
solid are therefore taken from the one condition in which the proposed
mechanism is absent.

Third, and most seriously, 7.8 THz appears to have been selected because
it is the frequency at which the model's own ignition criterion is
satisfied, rather than identified independently and then found to
satisfy it. A parameter chosen to make a criterion hold cannot
afterwards be cited as evidence that the criterion holds.

None of this shows that no suitable resonance exists. It shows that the
specific value is model-selected rather than measured, and that the
geometric and carrier-frequency results derived from it inherit that
status. Broadband terahertz spectroscopy of hydrated glutamate above 5
THz would settle the matter directly, which is why it appears in the
falsification list of §6.4.

\begin{center}\rule{0.5\linewidth}{0.5pt}\end{center}

\hypertarget{ethics}{%
\section*{Ethics}\label{ethics}}
\addcontentsline{toc}{section}{Ethics}

This work involved no human participants, no animal subjects and no
field data. No ethical approval was required.

\hypertarget{data-accessibility}{%
\section*{Data accessibility}\label{data-accessibility}}
\addcontentsline{toc}{section}{Data accessibility}

No new data were generated in the course of this study. Every quantity
reported is derived from published sources cited in the reference list,
or recomputed from those sources by the methods stated.

The reference implementation of the seven criteria, together with
scripts that regenerate every table in this paper and assert every
quoted value against its defining formula, is openly archived at Zenodo
under the MIT licence: https://doi.org/10.5281/zenodo.21837368 (Kopel
2026b). It comprises three Python files with no dependencies beyond the
standard library and requires no data to run.

An earlier and partial statement of the framework of §3 to §5, without
the case analysis of §2, is openly deposited at
https://doi.org/10.5281/zenodo.21837082 (Kopel 2026a).

\hypertarget{declaration-of-ai-use}{%
\section*{Declaration of AI use}\label{declaration-of-ai-use}}
\addcontentsline{toc}{section}{Declaration of AI use}

During the preparation of this work the author used Claude, an AI
assistant developed by Anthropic, for the following purposes: to search
for and retrieve candidate references and to verify their bibliographic
metadata against Crossref and PubMed; to recompute the quantitative
claims reported here from their defining formulae and to flag
discrepancies between reported and recomputed values; to write the
Python reference implementation archived at Kopel (2026b); and to draft
and revise manuscript text. After using this tool the author reviewed
and edited the content as needed and takes full responsibility for the
content of this publication.

\hypertarget{authors-contributions}{%
\section*{Authors' contributions}\label{authors-contributions}}
\addcontentsline{toc}{section}{Authors' contributions}

E.K.: conceptualization, formal analysis, investigation, methodology,
software, writing (original draft), writing (review and editing).

\hypertarget{conflict-of-interest-declaration}{%
\section*{Conflict of interest
declaration}\label{conflict-of-interest-declaration}}
\addcontentsline{toc}{section}{Conflict of interest declaration}

I declare I have no competing interests.

\hypertarget{funding}{%
\section*{Funding}\label{funding}}
\addcontentsline{toc}{section}{Funding}

This research received no specific grant from any funding agency in the
public, commercial or not-for-profit sectors.

\begin{center}\rule{0.5\linewidth}{0.5pt}\end{center}

\raggedright

\hypertarget{references}{%
\section*{References}\label{references}}
\addcontentsline{toc}{section}{References}

Attwell, D., Laughlin, S.B., 2001. An energy budget for signaling in the
grey matter of the brain. J. Cereb. Blood Flow Metab. 21, 1133--1145.
https://doi.org/10.1097/00004647-200110000-00001

Blanco-Duque, C., Bond, S.A., Krone, L.B., Dufour, J.-P., Gillen,
E.C.P., Purple, R.J., Kahn, M.C., Bannerman, D.M., Mann, E.O.,
Achermann, P., Olbrich, E., Vyazovskiy, V.V., 2024. Oscillatory-Quality
of sleep spindles links brain state with sleep regulation and function.
Sci. Adv. 10, eadn6247. https://doi.org/10.1126/sciadv.adn6247

Bolterauer, H., 1999. Elementary arguments that the Wu-Austin
Hamiltonian has no finite ground state. Bioelectrochem. Bioenerg. 48,
301.

Casanova, M.F., Shaban, M., Ghazal, M., El-Baz, A.S., Casanova, E.L.,
Sokhadze, E.M., 2021. Ringing decay of gamma oscillations and
transcranial magnetic stimulation therapy in autism spectrum disorder.
Appl. Psychophysiol. Biofeedback 46, 161--173.
https://doi.org/10.1007/s10484-021-09509-z

Cousijn, H., Haegens, S., Wallis, G., Near, J., Stokes, M.G., Harrison,
P.J., Nobre, A.C., 2014. Resting GABA and glutamate concentrations do
not predict visual gamma frequency or amplitude. Proc. Natl. Acad. Sci.
U.S.A. 111, 9301--9306. https://doi.org/10.1073/pnas.1321072111

Del Giudice, E., Doglia, S., Milani, M., Vitiello, G., 2005. Coherent
quantum electrodynamics in living matter. Electromagn. Biol. Med. 24,
199--210.

Keppler, J., 2023. Scrutinizing the feasibility of macroscopic quantum
coherence in the brain: a field-theoretical model of cortical dynamics.
bioRxiv 2023.03.03.530961.

Keppler, J., 2024. Laying the foundations for a theory of consciousness.
Front. Hum. Neurosci. https://doi.org/10.3389/fnhum.2024.1379191

Keppler, J., 2025. Macroscopic quantum effects in the brain. Front. Hum.
Neurosci. 19, 1676585. https://doi.org/10.3389/fnhum.2025.1676585

Kopel, E., 2026a. What must a biological structure be to carry
addressable information? Seven criteria, and a screen for candidate
carriers. Zenodo. https://doi.org/10.5281/zenodo.21837082

Kopel, E., 2026b. Criteria for biological information carriers:
reference implementation, v1.0.0. Zenodo.
https://doi.org/10.5281/zenodo.21837368

Keppler, J., 2026. Toward a true understanding of consciousness: the
explanatory power behind the non-physicalist paradigm. Front. Hum.
Neurosci. 20, 1815678. https://doi.org/10.3389/fnhum.2026.1815678

Korotkevich, A.A., Bakker, H.J., 2022. Ultrafast vibrational dynamics of
aqueous acetate and terephthalate. J. Chem. Phys. 156, 094501.
https://doi.org/10.1063/5.0082462

Kuroda, D.G., Hochstrasser, R.M., 2011. Two-dimensional infrared
spectral signature and hydration of the oxalate dianion. J. Chem. Phys.
135, 204502. https://doi.org/10.1063/1.3658461

Kuroda, D.G., Hochstrasser, R.M., 2012. Dynamic structures of aqueous
oxalate and the effects of counterions seen by 2D IR. Phys. Chem. Chem.
Phys. 14, 6219. https://doi.org/10.1039/c2cp23892f

Kuroda, D.G., Vorobyev, D.Yu., Hochstrasser, R.M., 2010. Ultrafast
relaxation and 2D IR of the aqueous trifluorocarboxylate ion. J. Chem.
Phys. 132, 044501. https://doi.org/10.1063/1.3285265

Lowet, E., Roberts, M.J., Peter, A., Gips, B., De Weerd, P., 2017. A
quantitative theory of gamma synchronization in macaque V1. eLife 6,
e26642. https://doi.org/10.7554/eLife.26642

Muthukumaraswamy, S.D., Edden, R.A.E., Jones, D.K., Swettenham, J.B.,
Singh, K.D., 2009. Resting GABA concentration predicts peak gamma
frequency and fMRI amplitude in response to visual stimulation in
humans. Proc. Natl. Acad. Sci. U.S.A. 106, 8356--8361.
https://doi.org/10.1073/pnas.0900728106

Pesnot Lerousseau, J., Trébuchon, A., Morillon, B., Schön, D., 2021.
Frequency selectivity of persistent cortical oscillatory responses to
auditory rhythmic stimulation. J. Neurosci. 41, 7991--8006.
https://doi.org/10.1523/JNEUROSCI.0213-21.2021

Preparata, G., 1995. QED Coherence in Matter. World Scientific,
Singapore.

Ray, S., Maunsell, J.H.R., 2010. Differences in gamma frequencies across
visual cortex restrict their possible use in computation. Neuron 67,
885--896.

Reimers, J.R., McKemmish, L.K., McKenzie, R.H., Mark, A.E., Hush, N.S.,
2009. Weak, strong, and coherent regimes of Fröhlich condensation and
their applications to terahertz medicine and quantum consciousness.
Proc. Natl. Acad. Sci. U.S.A. 106, 4219--4224.
https://doi.org/10.1073/pnas.0806273106

Rife, D., Boorstyn, R., 1974. Single tone parameter estimation from
discrete-time observations. IEEE Trans. Inf. Theory 20, 591--598.
https://doi.org/10.1109/TIT.1974.1055282

Spyropoulos, G., Saponati, M., Dowdall, J.R., Schölvinck, M.L., Bosman,
C.A., Lima, B., Peter, A., Onorato, I., Klon-Lipok, J., Roese, R.,
Neuenschwander, S., Fries, P., Vinck, M., 2022. Spontaneous variability
in gamma dynamics described by a damped harmonic oscillator driven by
noise. Nat. Commun. 13, 2019. https://doi.org/10.1038/s41467-022-29674-x

Tan, H.-R.M., Gross, J., Uhlhaas, P.J., 2016. MEG sensor and source
measures of visually induced gamma-band oscillations are highly
reliable. NeuroImage 137, 34--44.
https://doi.org/10.1016/j.neuroimage.2016.05.006

Tegmark, M., 2000. Importance of quantum decoherence in brain processes.
Phys. Rev.~E 61, 4194--4206.

Wu, T.M., Austin, S., 1977. Bose condensation in biosystems. Phys. Lett.
A 64, 151.

\end{document}